\documentclass[multicol]{elsart}

\usepackage{adnd}
\usepackage{longtable}

\usepackage{mathptmx}

\usepackage{amsmath}

\usepackage{amssymb}

\usepackage{graphicx}

\begin{document}

\begin{frontmatter}

\journal{Atomic Data and Nuclear Data Tables}

\copyrightholder{Elsevier Science}

\runtitle{Electric dipole polarizabilities}
\runauthor{Derevianko, Porsev, and Babb}


\title{Electric
dipole polarizabilities at imaginary frequencies 
for the alkali-metal,
alkaline-earth, and inert gas atoms
}

\author[UNLV]{Andrei Derevianko}
\ead{andrei@physics.unr.edu}
\and         
\author[UNLV,PNPI]{Sergey G. Porsev}
\ead{sporsev@gmail.com}
\and         
\author[ITAMP]{James F. Babb}
\address[UNLV]{Physics Department, Univ. of Nevada, Reno, NV 89557-0058}
\address[PNPI]{Petersburg Nuclear Physics Institute, Gatchina, Leningrad District, 188300, Russia}
\address[ITAMP]{ITAMP, Harvard-Smithsonian Center for Astrophysics
\\60 Garden St., Cambridge, MA 02138}
\ead{jbabb@cfa.harvard.edu}

\date{Feb 23, 2009} 

\begin{abstract}
  The electric dipole polarizabilities evaluated at imaginary
  frequencies for hydrogen, the alkali-metal atoms, the alkaline earth atoms,
  and the inert gases are tabulated along with 
  the resulting values of the  atomic static
  polarizabilities,  the atom-surface interaction constants, and the
  dispersion (or van der Waals)
  constants for the homonuclear and the heteronuclear diatomic
  combinations of the atoms.
\end{abstract}

\end{frontmatter}





\newpage
\tableofcontents
\listofDtables

\vskip5pc


\section{Introduction}

Dynamic electric dipole polarizability functions describe the response
of atoms to applied oscillating and fluctuating electric fields and
consequently they are necessary ingredients for many
applications~\cite{BonKre97}.  When evaluated as functions of
imaginary frequencies they can be used for
straightforward calculations of long-range interactions such as the
dispersion (or van der Waals) and retarded (or Casimir-Polder)
potentials between two atoms, the potential between an atom and a
surface~\cite{CasPol48} or of an atom between two
surfaces~\cite{Bar87a,ZhoSpr95}, and the dispersion (or
Axilrod-Teller) potential between three
atoms~\cite{ChaDal65b,DalDav66,Goo73}.  Other applications include
evaluation of the Lifshitz formula for the free energy of macroscopic
media at zero and non-zero temperature~\cite{AntPitStr08} and quantum
reflection studies~\cite{ArnFriMad07}.  Numerical values of dynamic
polarizabilities are useful tests for computations of atomic structure
using wave function based methods~\cite{DerJohSaf99} or density
functional theory methods~\cite{LimCal05,ChuDalGro07,AloMan07}.

We have carried out a program of calculating the dynamic electric
dipole polarizabilities of the alkali-metal atoms and of the
alkaline-earth atoms for use in determining long-range dispersion
coefficients~\cite{DerJohSaf99,DerBabDal01,PorDer06Z,PorDer02} for
photoassociation and ultra-cold atom scattering studies.  For those
papers it was not necessary to list the values of the polarizabilities
as functions of imaginary frequencies.  
Since they are useful in a range of contexts, we present 
herein a set of values in readily usable form.
For completeness, we also tabulate
the corresponding static polarizabilities~\cite{TeaPac71,Mil08}, dispersion
coefficients, and atom-surface interaction coefficients.

\section{The atomic dynamic electric dipole polarizability}

The atomic dynamic electric dipole polarizability is 
given by the
expression
\begin{equation}
\label{Eqn_alpha}
\alpha( u) = \mbox{\Large{\sf{}S}}'_n \frac{f_n}{\omega_n^2 - u^2} \, ,
\end{equation}
with $f_n$  the absorption
oscillator strengths, $\omega_n$  the excitation energies,
$u$ the frequency of the applied electric field,
and
$\mbox{\sf{}S}'$  a combined summation over final discrete
states and integration over final continuum states 
excluding the initial state.
The function
$\alpha (i\omega)$
is defined by the replacement of
$u$ by $i\omega$ in Eq.~(\ref{Eqn_alpha})~\cite{ChaDal65b,HirBroEps64,DalDav66}
and it 
can be constructed 
from calculated bound and continuum properties,
or variationally, or obtained
from fits to combinations
of theoretical and  empirical data.
The function $\alpha (i\omega)$ is real and smooth.
Oscillator strength sum rules dictate that it vary  from
the value $\alpha (0)$ at $\omega=0$ to 
the value $N_e/\omega^2$ as $\omega\sim\infty$,
where $N_e$ is the number of electrons in the atom.

The dispersion (or van der Waals) constant $C_6(AB)$ enters in the
interaction potential $-{C_6(AB)}/{R^6}$ between two ground state
atoms $A$ and
$B$ at large internuclear distances $R$, see for example,
Ref.~\cite{DalDav66}.  The constant $C_6 (AB)$ can be expressed
as 
\begin{equation}
\label{C6}
C_6(AB) = \frac{3}{\pi}
\int_0^\infty \,d\omega \alpha^A(i \omega) \alpha^B(i \omega)  \, ,
\end{equation}
where $\alpha^A (i\omega)$ and $\alpha^B (i\omega)$, respectively, correspond
to atom $A$ and atom $B$.

The potential between an atom  and a perfectly conducting metal surface
is given in terms of the normal distance $z$ between the atom and the  wall 
as $-C_3/z^3$, where the atom-surface  interaction coefficient is
\begin{equation}
\label{C3}
C_3  = \frac{1}{4\pi} \int_0^\infty \,d\omega \alpha (i\omega)\, .
\end{equation}
Integration of Eq.~(\ref{C3}) for $C_3$ yields an alternate form in terms
of the ground state atomic wave function $|0\rangle$,
\begin{equation}
\label{C3-direct}
C_3  = \frac{1}{12} 
   \left \langle 0 \left| \left(\sum_{i=1}^{N_e}{\mathbf r}_i \right)^2
        \right| 0 \right\rangle ,
\end{equation}
where  ${\mathbf r}_i$ is  the position vector
of electron $i$ from the nucleus.

\section{Sources of polarizability data}

\subsection{Alkali metals}

The dynamic polarizability for hydrogen is known
analytically~\cite{KarKol63,Gav67}.  It is also available in
parameterized forms using pseudo dipole oscillator strength
distributions (or ``pseudo-DOSDs'')~\cite{JohEpsMea67} as well as in
tabulations at various imaginary frequencies~\cite{BisPip92}.  For the
present work, it was calculated to sufficient accuracy using direct
summation over a relativistic B-spline basis for the Coulomb field.

For alkali-metal atoms we employ dynamic polarizabilities calculated
previously in Ref.~\cite{DerJohSaf99,DerBabDal01,GeuJeuDer01}. In that
work, dynamic polarizabilities were obtained by combining
high-precision experimental data for matrix elements of principal
transitions with several many-body methods of various accuracy.  In
particular, the relativistic linearized coupled-cluster method
limited to single and double (SD) excitations from the reference state was
used for calculation of oscillator strengths for the  first several excited
states. The results for SD matrix elements were presented
in~\cite{SafJohDer99}. The relativistic random-phase approximation
(RRPA) was employed for calculation of the contributions to the
polarizabilities from core excitations, and, finally, the
Dirac-Hartree-Fock (DHF) method was used to obtain less significant
contributions.  

The resulting static polarizabilities and dynamic polarizabilities 
at imaginary frequencies for hydrogen and the alkali-metal atoms
are collected in Table~\ref{Tab:AlphaAlk}.

\subsection{Alkaline-earth metals}

To calculate the dynamic polarizabilities for
alkaline-earth metals dimers we employed several atomic relativistic
many-body methods of varying accuracy~\cite{PorDer06Z,PorDer02}.  The
intermediate states in the sum appearing in Eq.~(\ref{Eqn_alpha}) were
formally 
separated into valence  and core-excited states 
giving
\begin{equation}
\alpha ( i \omega ) = \alpha_v ( i \omega ) + \alpha_c ( i \omega ) +
\alpha_{cv} ( i \omega ) \, .
\label{Eqn_alphaBreak}
\end{equation}
The valence contribution $\alpha_v$, which gives the dominant
contribution to $C_6$, was evaluated with the relativistic
configuration interaction (CI) method coupled with many-body
perturbation theory (MBPT) \cite{DzuFlaKoz96b,DzuKozPor98}.  The smaller
contributions of core-excited states $\alpha_c$ were estimated using
RRPA for the atomic core. In this method excitations of core electrons
are allowed into the occupied valence shell and we introduced the
correction $\alpha_{cv}$ to remove these
Pauli-exclusion-principle-violating excitations; this small correction
was evaluated using the DHF method.  The values of polarizabilities
were further adjusted using accurate theoretical and experimental data
for the electric-dipole matrix elements and the energies of 
the principal
transitions (see \cite{PorDer06Z,PorDer02} for details).  The
resulting static polarizabilities and dynamic polarizabilities at
imaginary frequencies for the alkaline-earth
atoms are collected in Table~\ref{Tab:AlphaAE}.

\subsection{Inert gases}

Dynamic polarizability data for the helium atom were obtained in the
present work using an essentially exact relativistic CI method.

For the dynamic polarizabilities of Ne, Ar, Kr, and Xe, we used the
semi-empirical dipole oscillator strength distributions (DOSD)
constructed by Kumar and Meath~\cite{KumMea85} using constraints such
as oscillator strength sum rules applied to certain available
experimental data.  The values in Table~\ref{Tab:AlphaNob} constitute a
different representation of the data presented by Kumar and
Meath~\cite{KumMea85}.

The resulting polarizabilities for the inert gases atoms are collected
in Table~\ref{Tab:AlphaNob}.

\section{The tabulated data}

The values of the static polarizability, $\alpha (0)$ are given in the
first line of Tables~\ref{Tab:AlphaAlk}--\ref{Tab:AlphaNob}.
Comparisons of values of $\alpha (0)$ from various sources for various
atoms can be found in the literature, cf.~\cite{TeaPac71,Mil08}.

A typical application of dynamic polarizabilities involves integration
of a product of $\alpha(i \omega)$ with a smoothly behaving function
$f(\omega)$, $I = \int^\infty_0 f(\omega) \alpha(i \omega) d \omega$.
The integration over frequencies ranges from zero to infinite values.
An economic and accurate method for evaluating such integrals is the
method of Gaussian quadratures~\cite{BisPip92}.  In this method the
above integration is replaced by a finite sum $I = \sum_{k=1}^{N_g}
w_k \, f(\omega_k) \alpha(i \omega_k ) $ over values of $f(\omega_k)
\alpha(i \omega_k )$ tabulated at certain frequencies $\omega_k$
yielding an $N_g$-point quadrature, where 
each term in the sum is weighted by factors $w_k$. In this work we use
points and weights listed in Table~\ref{Tab:GQ} and $N_g=50$.

\subsection{Applications to atom-atom dispersion coefficients}

The dispersion coefficients are computed
by evaluating Eq.~(\ref{C6}) using the Gaussian quadrature method 
yielding 
\begin{eqnarray}
\label{C6_Gauss}
C_6(AB) &=& \frac{3}{ \pi}  \,\sum_{k=1}^{50} w_k \, \alpha^A(i \omega_k )
     \, \alpha^B(i \omega_k ) \, .
\end{eqnarray}
For comparisons listing various previous determinations of the van der
Waals coefficients, see Maeder and Kutzelnigg~\cite{MaeKut79}, and our
previous papers~\cite{DerJohSaf99,PorDer06Z}.  The resulting
coefficients for various pairs of alkali-metal, alkaline-earth, and
noble-gas atoms are listed in
Tables~\ref{Tab:C6Alk}--\ref{Tab:C6NobxAlk}.

For the alkali-metals, the accuracy of calculation of $C_6$ dispersion
coefficients using Eq.~(\ref{C6_Gauss}) and presented in
Table~\ref{Tab:C6Alk} was estimated using \textit{i}) experimental
error bars for principal transitions and \textit{ii})) by comparison
of the atom-wall interaction constant $C_3$ obtained as an integral of
the dynamic polarizability, Eq.~(\ref{C3}), with the SD {\em ab
  initio} calculation of Eq.~(\ref{C3-direct})~\cite{DerJohFri98}.
The estimated accuracy for the $C_6$ coefficients was at the level of
0.1\% for Li, with less accurate results for heavier systems, reaching
1\% for Cs and 1.5\% for Fr.  Agreement with subsequently determined
experimental values for $C_6$ coefficients was excellent. For example,
for Cs, the predicted value was found to be in an excellent agreement
with the results from Feshbach resonance spectroscopy with ultracold
atoms~\cite{DerBabDal01}.  Similarly, in atomic units, our value 399.8
for the static polarizability of Cs listed in Table~\ref{Tab:AlphaAlk}
is in 0.3\% agreement with the experimental value~\cite{AmiGou03} of
401.0(6).

For the alkaline-earth metals, in Table~\ref{Tab:C6AE} we present
$C_6$ coefficients for homonuclear and heteronuclear dimers. 
The estimates
of the uncertainties in these coefficients was discussed in detail
in Ref.~\cite{PorDer02}.

For the inert gases, in Table~\ref{Tab:C6Nob},
we have indicated uncertainties of 1\% for the tabulated
values of $C_6$, as suggested by Kumar and Meath~\cite{KumMea85}.
The analysis of the latest experimental data for inert gases by
Bulanin and Kislyakov~\cite{BulKis98} is in excellent agreement with
the results of Ref.~\cite{KumMea85} for values of the dynamic
polarizability $\alpha(i\omega)$ below the first resonance.  However,
there are  differences at the 1\%~level in the values of 
oscillator strength sums obtained in 
Ref.~\cite{KumMea85} and Ref.~\cite{BulKis98}, supporting 
the estimate of a 1\%~level of accuracy for the  $C_6$ coefficients
obtained by Kumar and Meath~\cite{KumMea85}.

For the heteronuclear atom pairs, the dispersion coefficient $C_6(AB)$
of the alkali-metals and the alkaline-earths are listed in
Table~\ref{Tab:C6AlkxAE}, the values for the alkaline-earths and the
noble gases are listed in Table~\ref{Tab:C6AExNob}, and those for the
alkali-metals and the noble gases are listed in
Table~\ref{Tab:C6NobxAlk}.

The uncertainty  $dC_6 (AB)$ 
in $C_6(AB)$ 
was estimated using the expression
\begin{equation}
 dC_6(AB)/C_6(AB) = 
 {\frac{1}{2}} 
 [ (dC_6(AA) /C_6(AA))^2 + (dC_6(BB) /C_6(BB))^2 ]^{1/2} \,.
\end{equation}

\subsection{Applications to atom-surface interactions}

The atom-surface interaction 
coefficients 
were computed with Eq.~(\ref{C3}) using the Gaussian quadrature
method yielding 
\begin{eqnarray}
\label{C3_Gauss}
  C_3 &=& \frac{1}{4 \pi} \,\sum_{k=1}^{50} w_k \, \alpha(i \omega_k ) \, .
\end{eqnarray}
The resulting values  are given in Table~\ref{Tab:C3}.


\begin{appendix}

\section{GAUSS-LEGENDRE QUADRATURE}
We determine the tabulated grid points and weights in two steps.
First we obtain the Gauss-Legendre abscissas $x_k$ and weights $g_k$
defined on the interval $(0,1)$ with  $N_g=50$ points. The weights and
abscissas are generated by routine \texttt{gauleg} of
Ref.~\cite{PreFlaTeu92}.  Further we use mapping function $\omega_k =
2 \tan(\pi/2 x_k )$, so that the resulting values of $\omega$ sample
the entire integration range. The Gauss-Legendre weights are also
properly redefined to incorporate the Jacobian of the coordinate
transformation, $w_k = g_k \pi /\cos(\pi/2 x_k)^2 $.
The final values are given in Table~\ref{Tab:GQ}.

\begin{table}[h]
\caption{Gaussian quadrature abscissas $\omega_k$ and weights $w_k$
for a 50-pair integration, with $k=1,...,50$.
\label{Tab:GQ} }
\begin{center}
\begin{tabular}{lllllll}
\hline \hline
\multicolumn{1}{c}{$k$}&\multicolumn{1}{c}{$\omega_k$}&
                      \multicolumn{1}{c}{$w_k$} &
   \multicolumn{1}{c}{ } &
                          \multicolumn{1}{c}{$k$}&
   \multicolumn{1}{c}{$\omega_k$}&\multicolumn{1}{c}{$w_k$}\\
\hline
  1  &  0.00178065  &  0.00456886  &&  26  &  2.10016  &  0.205361\\
  2  &  0.00937463  &  0.0106185   &&  27  &  2.31644  &  0.227799\\
  3  &  0.0230068  &  0.0166378    &&  28  &  2.55713  &  0.254343\\
  4  &  0.042631  &  0.0225996     &&  29  &  2.82683  &  0.286005\\
  5  &  0.0681817  &  0.0284889    &&  30  &  3.13128  &  0.324106\\
  6  &  0.0995817  &  0.0342971    &&  31  &  3.47776  &  0.370381\\
  7  &  0.136748  &  0.0400227     &&  32  &  3.87553  &  0.427154\\
  8  &  0.179602  &  0.0456718     &&  33  &  4.3366  &  0.497572\\
  9  &  0.228071  &  0.0512592     &&  34  &  4.87665  &  0.585976\\
  10  &  0.282107  &  0.0568084    &&  35  &  5.51655  &  0.698465\\
  11  &  0.341686  &  0.0623519    &&  36  &  6.28448  &  0.843785\\
  12  &  0.406823  &  0.0679318    &&  37  &  7.21927  &  1.03477\\
  13  &  0.477579  &  0.0735998    &&  38  &  8.37558  &  1.29076\\
  14  &  0.554072  &  0.0794176    &&  39  &  9.83228  &  1.64181\\
  15  &  0.636488  &  0.085458     &&  40  &  11.7066  &  2.13626\\
  16  &  0.725091  &  0.0918056    &&  41  &  14.179  &  2.85525\\
  17  &  0.820235  &  0.0985589    &&  42  &  17.5384  &  3.94176\\
  18  &  0.922382  &  0.105832     &&  43  &  22.2715  &  5.66354\\
  19  &  1.03212  &  0.113758      &&  44  &  29.2508  &  8.56095\\
  20  &  1.15017  &  0.122493      &&  45  &  40.168  &  13.8343\\
  21  &  1.27743  &  0.132222      &&  46  &  58.6667  &  24.5132\\
  22  &  1.41501  &  0.143164      &&  47  &  93.8285  &  49.7405\\
  23  &  1.56425  &  0.155587      &&  48  &  173.862  &  125.732\\
  24  &  1.72679  &  0.169813      &&  49  &  426.684  &  483.298\\
  25  &  1.90461  &  0.18624       &&  50  &  2246.37  &  5763.83\\
\hline \hline
\end{tabular}
\end{center}
\end{table}




\end{appendix}

\ack
Supported in part by the NSF through a grant for the Institute of
Theoretical Atomic, Molecular and Optical Physics (ITAMP) at Harvard
University and Smithsonian Astrophysical Observatory. AD was supported
in part by the NSF, and SGP by the Russian Foundation for Basic
Research under Grants No. 07-02-00210-a and No. 08-02-00460-a.



\bibliographystyle{aip}
\bibliography{all,allsup}

\newpage



\begin{table}[h]
\caption{
Dispersion coefficients $C_6$
for alkali-metal atom pairs in atomic units.
\label{Tab:C6Alk} }
\begin{center}
\begin{tabular}{llllllll}
\hline \hline
      &  H      &    Li    &     Na    &   K        &  Rb      &    Cs      &   Fr       \\
\hline
  H   &  6.498  &  66.36(5)&  73.83(9) &  111.2(2)  &  124.3(3)&  149.7(8)  &  142.3(1.2)\\
  Li  &         &  1389(2) &  1467(2)  &  2322(5)   &  2545(7) &  3065(17)  &  2682(23)\\
  Na  &         &          &  1556(4)  &  2447(6)   &  2683(7) &  3227(18)  &  2842(24)\\
  K   &         &          &           &  3897(15)  &  4274(13)&  5159(30)  &  4500(39)\\
  Rb  &         &          &           &            &  4690(23)&  5663(34)  &  4946(44)\\
  Cs  &         &          &           &            &          &  6846(74)  &  5968(60)\\
  Fr  &         &          &           &            &          &            &  5256(89)\\
\hline \hline
\end{tabular}
\end{center}
\end{table}

\begin{table}[h]
\caption{
Dispersion coefficients $C_6$ and their estimated uncertainties
(in parentheses)
for alkaline-earth metal atom pairs in atomic units.
\label{Tab:C6AE} }
\begin{center}
\begin{tabular}{llllllll}
\hline \hline
      &  Be  &  Mg  &  Ca  &  Sr  &  Ba\\
\hline
  Be  &  214(3)     &    364(4)       &    652(7)  &  782(6)     &  992(9)  \\
  Mg  &               &  627(12)      &  1138(14)  &  1369(13)   &  1742(21)\\
  Ca  &               &               &  2121(35)  &  2564(21)   &  3294(36)\\
  Sr  &               &               &            &  3103(7)    &  3994(29)\\
  Ba  &               &               &            &             &  5160(74)\\
\hline \hline
\end{tabular}
\end{center}
\end{table}

\begin{table}[h]
\caption{
Dispersion coefficients $C_6$
and their estimated uncertainties
(in parentheses)
for noble-gas atom pairs in atomic units.
\label{Tab:C6Nob} }
\begin{center}
\begin{tabular}{llllllll}
\hline \hline
     &  He  &  Ne  &  Ar  &  Kr  &  Xe\\
\hline
  He  &  1.461 &  3.03(2)  &  9.55(5)  &  13.42(7) &  19.6(1)\\
  Ne  &        &  6.38(6)  &  19.5(1)  &  27.3(2)  &  39.7(3)\\
  Ar  &        &           &  64.3(6)  &  91.1(6)  &  134.5(9)\\
  Kr  &        &           &           &  130(1)   &  192(2) \\
  Xe  &        &           &           &           &  286(3) \\
\hline \hline
\end{tabular}
\end{center}
\end{table}

\begin{table}[h]
\caption{
Dispersion coefficients $C_6$
and their estimated uncertainties
(in parentheses)
for alkali-metal--alkaline-earth metal atom pairs in atomic units.
\label{Tab:C6AlkxAE} }
\begin{center}
\begin{tabular}{llllllll}
\hline \hline
      &  Be          &  Mg        &  Ca        &  Sr        &  Ba\\
\hline
  H   &  34.8(2)     &  57.4(6)   &  98.3(8)   &  117.5(1)  &  148(1)\\
  Li  &  478(3)      &  853(8)    &  1660(14)  &  2022(3)   &  2637(19)\\
  Na  &  521(4)      &  926(9)    &  1782(15)  &  2167(4)   &  2815(21)\\
  K   &  790(6)      &  1411(14)  &  2756(23)  &  3362(8)   &  4396(33)\\
  Rb  &  873(7)      &  1556(15)  &  3030(26)  &  3697(10)  &  4832(37)\\
  Cs  &  1045(9)     &  1863(20)  &  3635(36)  &  4437(24)  &  5809(52)\\
  Fr  &  963(11)     &  1701(22)  &  3265(39)  &  3974(34)  &  5170(57)\\
\hline \hline
\end{tabular}
\end{center}
\end{table}


\begin{table}[h]
\caption{Dispersion coefficients $C_6$ and their estimated uncertainties
(in parentheses) for alkaline-earth -- noble gas pairs in atomic units.
\label{Tab:C6AExNob} }
\begin{center}
\begin{tabular}{llllllll}
\hline \hline
     &  He  &  Ne  &  Ar  &  Kr  &  Xe\\
\hline
  Be  &  13.23(9)      &  26.0(2)    &  97.9(8)  &   143(1)  &  221(2)\\
  Mg  &  21.3(2)       &  41.9(5)  &  159(2)  &  234(3)  &  363(4)\\
  Ca  &  35.7(3)       &  70.2(7)  &  268(3)  &  396(4)  &  617(6)\\
  Sr  &  42.84(5)      &  84.2(4)  &  321(2)  &  474(2)  &  739(4)\\
  Ba  &  53.7(4)       &  105.7(9)   &  403(4)  &  594(5)  &  927(8)\\
\hline \hline
\end{tabular}
\end{center}
\end{table}


\begin{table}[h]
\caption{Dispersion coefficients $C_6$ and their estimated uncertainties
(in parentheses) for alkali-metal -- noble gas pairs in atomic units.
\label{Tab:C6NobxAlk} }
\begin{center}
\begin{tabular}{llllllll}
\hline \hline
     &  He  &  Ne  &  Ar  &  Kr  &  Xe\\
\hline
  H   &  2.821     &  5.64(3)  &  19.8(1)   &  28.5(2) &  42.8(2)\\
  Li  &  22.44(2)  &  43.6(2)  &  173(1)    &  259(1)  &  409(2)\\
  Na  &  25.51(3)  &  49.8(3)  &  195(1)    &  291(2)  &  458(2)\\
  K   &  38.86(8)  &  76.1(4)  &  296(2)    &  440(3)  &  692(4)\\
  Rb  &  44.07(11) &  86.5(5)  &  334(2)    &  494(3)  &  776(4)\\
  Cs  &  53.6(3)   &  105.4(8)   &  404(3)    &  598(4)  &  936(7)\\
  Fr  &  52.4(4)   &  103(1)     &  390(4)    &  575(6)  &  896(9)\\
\hline \hline
\end{tabular}
\end{center}
\end{table}


\begin{table}[h]
\caption{Coefficients $C_3$ for the atom-surface interaction
in atomic units.
\label{Tab:C3} }
\begin{center}
\begin{tabular}{llllllll}
\hline \hline
\multicolumn{2}{c}{Group IA}&\multicolumn{1}{c}{~}&
\multicolumn{2}{c}{Group IIA}&\multicolumn{1}{c}{~}&
\multicolumn{2}{c}{Group VIII}\\
  Atom& $C_3$  &&  Atom  &  $C_3$ &&   Atom & $C_3$\\
\hline            
  H   & 0.25   &&        &        &&   He   & 0.1881 \\
  Li  & 1.512  &&  Be    &  1.01  &&   Ne   & 0.4751 \\
  Na  & 1.871  &&  Mg    &  1.666 &&   Ar   & 1.096 \\
  K   & 2.896  &&  Ca    &  2.744 &&   Kr   & 1.542 \\
  Rb  & 3.426  &&  Sr    &  3.382 &&   Xe   & 2.164 \\
  Cs  & 4.268  &&  Ba    &  4.293 &&        &       \\
  Fr  & 4.437  &&        &        &&        &       \\
\hline \hline
\end{tabular}
\end{center}
\end{table}





\clearpage
\newpage

\section*{EXPLANATION OF TABLES}\label{sec.eot}
\addcontentsline{toc}{section}{EXPLANATION OF TABLES}

\textbf{Table~\ref{Tab:AlphaAlk}:}~Dynamic
electric-dipole polarizabilities of hydrogen and alkali-metal atoms
at imaginary frequencies for use with the 50-point Gauss-Legendre quadrature
formula

\begin{tabular}{ll}
Column 1:     &Frequency: The value  $\omega=0$,  yielding the static polarizability  $\alpha (0)$ \\
              &The value $\omega_k$,  corresponding 
to the index $k$ in Table~\ref{Tab:GQ}\\
Columns 2--8: &Value of $\alpha(0)$ or $\alpha (i\omega_k)$ for the chemical element listed\\

\end{tabular}

\textbf{Table~\ref{Tab:AlphaAE}:}~Dynamic
electric-dipole polarizabilities of
alkaline-earth metal atoms
at imaginary frequencies for use with the 50-point Gauss-Legendre quadrature
formula

\begin{tabular}{ll}
Column 1:     &Frequency: The value  $\omega=0$, yielding the static polarizability  $\alpha (0)$ \\
              &The value $\omega_k$, corresponding 
               to the index $k$ in Table~\ref{Tab:GQ}\\
Columns 2--6: &Value of $\alpha(0)$ or $\alpha (i\omega_k)$ for the chemical element listed\\
\end{tabular}

\textbf{Table~\ref{Tab:AlphaNob}:}~Dynamic
electric-dipole polarizabilities of noble gas atoms
at imaginary frequencies for use with the 50-point Gauss-Legendre quadrature
formula

\begin{tabular}{l}
Same as Table~\ref{Tab:AlphaAE}
\end{tabular}



%
%






%

\newpage
\datatables

\setlength{\LTleft}{0pt}
\setlength{\LTright}{0pt} 
\footnotesize 
\begin{longtable}{@{\extracolsep\fill}llllllll}
\caption{Dynamic
electric-dipole polarizabilities of hydrogen and alkali-metal atoms
at imaginary frequencies for use with the 50-point Gauss-Legendre quadrature
method.\label{Tab:AlphaAlk}}\\[-8pt]
  \caption*{\small{See page \pageref{sec.eot} for Explanation of Tables}}\\
\endhead\\
\hline
 \multicolumn{1}{l}{$\omega$} & \multicolumn{1}{l}{H} & 
  \multicolumn{1}{l}{Li}  &  \multicolumn{1}{l}{Na}  &
  \multicolumn{1}{l}{K}  &  \multicolumn{1}{l}{Rb}  &
  \multicolumn{1}{l}{Cs}  &  \multicolumn{1}{l}{Fr}\\
\hline
  0           &   4.5     &  164.        &  162.6       &  290.2       &  318.6  &  399.8  &  317.8\\
\multicolumn{8}{c}{~}\\
$\omega_{1}$  &   4.4997  &   1.6386[2]  &    1.625[2]  &   2.8996[2]  &   3.1832[2]  &   3.9938[2]  &   3.1752[2]\\
$\omega_{2}$  &   4.4974  &   1.6094[2]  &   1.6025[2]  &   2.8329[2]  &   3.1076[2]  &     3.88[2]  &   3.1082[2]\\
$\omega_{3}$  &   4.4857  &   1.4727[2]  &   1.4948[2]  &   2.5306[2]  &   2.7667[2]  &   3.3819[2]  &   2.8035[2]\\
$\omega_{4}$  &    4.452  &    1.181[2]  &   1.2503[2]  &    1.934[2]  &   2.1035[2]  &   2.4771[2]  &   2.1973[2]\\
$\omega_{5}$  &   4.3798  &   8.2483[1]  &   9.2082[1]  &   1.2835[2]  &   1.3939[2]  &   1.5921[2]  &   1.5244[2]\\
$\omega_{6}$  &    4.252  &   5.3094[1]  &   6.1986[1]  &   8.0193[1]  &   8.7721[1]  &   9.9455[1]  &   1.0137[2]\\
$\omega_{7}$  &   4.0561  &    3.355[1]  &   4.0336[1]  &     5.06[1]  &    5.627[1]  &   6.4749[1]  &   6.9056[1]\\
$\omega_{8}$  &   3.7895  &   2.1584[1]  &   2.6414[1]  &   3.3376[1]  &    3.802[1]  &   4.5037[1]  &   4.9579[1]\\
$\omega_{9}$  &   3.4624  &   1.4337[1]  &   1.7756[1]  &    2.326[1]  &   2.7268[1]  &    3.343[1]  &   3.7583[1]\\
$\omega_{10}$  &   3.0959  &   9.8574  &   1.2345[1]  &   1.7106[1]  &    2.066[1]  &   2.6166[1]  &   2.9784[1]\\
$\omega_{11}$  &   2.7157  &   7.0022  &   8.8866  &   1.3188[1]  &   1.6373[1]  &   2.1283[1]  &   2.4368[1]\\
$\omega_{12}$  &   2.3453  &   5.1216  &   6.6142  &   1.0567[1]  &   1.3424[1]  &    1.776[1]  &   2.0363[1]\\
$\omega_{13}$  &   2.0018  &    3.844  &   5.0767  &   8.7219  &   1.1272[1]  &   1.5058[1]  &   1.7246[1]\\
$\omega_{14}$  &   1.6948  &   2.9512  &   4.0061  &   7.3564  &   9.6174  &   1.2886[1]  &   1.4729[1]\\
$\omega_{15}$  &   1.4274  &   2.3111  &   3.2398  &   6.2984  &   8.2871  &   1.1086[1]  &   1.2648[1]\\
$\omega_{16}$  &   1.1985  &   1.8417  &   2.6767  &   5.4454  &   7.1818  &   9.5655  &   1.0902[1]\\
$\omega_{17}$  &   1.0049  &   1.4902  &   2.2523  &   4.7358  &   6.2421  &   8.2676  &   9.4258\\
$\omega_{18}$  &   8.4223[-1]  &   1.2221  &   1.9245  &   4.1313  &   5.4316  &   7.1542  &     8.17\\
$\omega_{19}$  &     7.06[-1]  &   1.0139  &   1.6651  &   3.6081  &   4.7265  &   6.1966  &    7.098\\
$\omega_{20}$  &   5.9205[-1]  &   8.4974[-1]  &   1.4553  &   3.1504  &   4.1101  &    5.372  &     6.18\\
$\omega_{21}$  &   4.9668[-1]  &   7.1822[-1]  &   1.2819  &   2.7476  &   3.5702  &   4.6611  &   5.3912\\
$\omega_{22}$  &   4.1676[-1]  &   6.1136[-1]  &   1.1356  &   2.3919  &    3.097  &   4.0476  &    4.711\\
$\omega_{23}$  &   3.4963[-1]  &   5.2334[-1]  &   1.0098  &   2.0774  &   2.6821  &   3.5171  &   4.1221\\
$\omega_{24}$  &   2.9313[-1]  &   4.4989[-1]  &   8.9998[-1]  &   1.7993  &   2.3186  &   3.0576  &     3.61\\
$\omega_{25}$  &   2.4546[-1]  &   3.8785[-1]  &   8.0261[-1]  &   1.5535  &   2.0004  &   2.6583  &   3.1623\\
$\omega_{26}$  &   2.0515[-1]  &   3.3485[-1]  &   7.1525[-1]  &   1.3367  &    1.722  &   2.3104  &   2.7689\\
$\omega_{27}$  &   1.7102[-1]  &    2.891[-1]  &   6.3608[-1]  &   1.1456  &   1.4788  &   2.0061  &   2.4215\\
$\omega_{28}$  &   1.4207[-1]  &   2.4925[-1]  &   5.6376[-1]  &   9.7766[-1]  &   1.2664  &    1.739  &    2.113\\
$\omega_{29}$  &   1.1751[-1]  &   2.1428[-1]  &   4.9735[-1]  &   8.3028[-1]  &   1.0812  &   1.5036  &   1.8379\\
$\omega_{30}$  &   9.6682[-2]  &    1.834[-1]  &   4.3614[-1]  &    7.013[-1]  &   9.1991[-1]  &   1.2953  &   1.5914\\
$\omega_{31}$  &   7.9029[-2]  &   1.5602[-1]  &   3.7966[-1]  &   5.8879[-1]  &   7.7965[-1]  &   1.1104  &   1.3701\\
$\omega_{32}$  &   6.4103[-2]  &   1.3171[-1]  &   3.2757[-1]  &   4.9097[-1]  &   6.5784[-1]  &   9.4575[-1]  &   1.1712\\
$\omega_{33}$  &   5.1524[-2]  &   1.1013[-1]  &   2.7968[-1]  &   4.0628[-1]  &    5.522[-1]  &   7.9902[-1]  &   9.9237[-1]\\
$\omega_{34}$  &   4.0971[-2]  &   9.1031[-2]  &   2.3587[-1]  &    3.333[-1]  &   4.6071[-1]  &    6.683[-1]  &   8.3219[-1]\\
$\omega_{35}$  &   3.2172[-2]  &   7.4209[-2]  &   1.9607[-1]  &   2.7077[-1]  &   3.8156[-1]  &   5.5221[-1]  &   6.8947[-1]\\
$\omega_{36}$  &   2.4893[-2]  &   5.9517[-2]  &   1.6028[-1]  &   2.1753[-1]  &   3.1318[-1]  &   4.4973[-1]  &   5.6332[-1]\\
$\omega_{37}$  &   1.8931[-2]  &   4.6826[-2]  &   1.2848[-1]  &   1.7253[-1]  &   2.5416[-1]  &   3.6009[-1]  &   4.5304[-1]\\
$\omega_{38}$  &   1.4108[-2]  &   3.6019[-2]  &   1.0066[-1]  &   1.3479[-1]  &   2.0332[-1]  &    2.827[-1]  &    3.579[-1]\\
$\omega_{39}$  &   1.0263[-2]  &   2.6979[-2]  &   7.6781[-2]  &   1.0341[-1]  &   1.5966[-1]  &   2.1698[-1]  &   2.7711[-1]\\
$\omega_{40}$  &   7.2545[-3]  &   1.9581[-2]  &   5.6743[-2]  &   7.7582[-2]  &   1.2241[-1]  &   1.6227[-1]  &   2.0971[-1]\\
$\omega_{41}$  &   4.9534[-3]  &   1.3686[-2]  &   4.0389[-2]  &   5.6563[-2]  &   9.0954[-2]  &   1.1779[-1]  &   1.5452[-1]\\
$\omega_{42}$  &   3.2418[-3]  &   9.1389[-3]  &   2.7483[-2]  &   3.9715[-2]  &   6.4885[-2]  &   8.2549[-2]  &   1.1023[-1]\\
$\omega_{43}$  &   2.0123[-3]  &   5.7691[-3]  &   1.7703[-2]  &   2.6512[-2]  &   4.3891[-2]  &   5.5437[-2]  &   7.5407[-2]\\
$\omega_{44}$  &   1.1674[-3]  &   3.3923[-3]  &   1.0652[-2]  &   1.6525[-2]  &   2.7695[-2]  &   3.5233[-2]  &   4.8728[-2]\\
$\omega_{45}$  &   6.1937[-4]  &   1.8183[-3]  &   5.8682[-3]  &   9.3754[-3]  &   1.5939[-2]  &   2.0733[-2]  &   2.9029[-2]\\
$\omega_{46}$  &   2.9045[-4]  &   8.5884[-4]  &   2.8645[-3]  &   4.6648[-3]  &   8.0975[-3]  &   1.0867[-2]  &    1.535[-2]\\
$\omega_{47}$  &   1.1357[-4]  &   3.3735[-4]  &   1.1666[-3]  &    1.919[-3]  &   3.4329[-3]  &   4.7421[-3]  &   6.7792[-3]\\
$\omega_{48}$  &   3.3079[-5]  &   9.8504[-5]  &   3.5183[-4]  &   5.8509[-4]  &   1.0815[-3]  &   1.5266[-3]  &   2.2355[-3]\\
$\omega_{49}$  &   5.4925[-6]  &   1.6372[-5]  &   5.9695[-5]  &    1.012[-4]  &   1.9126[-4]  &   2.7665[-4]  &   4.1737[-4]\\
$\omega_{50}$  &   1.9816[-7]  &   5.9081[-7]  &   2.1693[-6]  &   3.7292[-6]  &   7.2051[-6]  &   1.0596[-5]  &   1.6446[-5]\\
\end{longtable}

\newpage

\setlength{\LTleft}{0pt}
\setlength{\LTright}{0pt} 
\footnotesize 
\begin{longtable}{@{\extracolsep\fill}llllll}
\caption{Dynamic 
electric-dipole polarizabilities of 
alkaline-earth metal atoms for use with the 50-point Gauss-Legendre quadrature
method.\label{Tab:AlphaAE} }\\[-8pt]
  \caption*{\small{See page \pageref{sec.eot} for Explanation of Tables}}\\
\endhead\\
\hline
\multicolumn{1}{l}{$\omega$}  & \multicolumn{1}{l}{Be}  &
      \multicolumn{1}{l}{Mg}  &  \multicolumn{1}{l}{Ca}  &
      \multicolumn{1}{l}{Sr}  &  \multicolumn{1}{l}{Ba}\\
\hline
0  &  37.76  &  71.26  &  157.1  &  197.2  &  273.5\\ 
\multicolumn{6}{c}{~}\\
$\omega_{1}$ &   3.7761[1]  &    7.126[1]  &   1.5713[2]  &   1.9717[2]  &   2.7346[2]\\ 
$\omega_{2}$ &   3.7678[1]  &   7.1031[1]  &   1.5605[2]  &   1.9555[2]  &   2.7034[2]\\ 
$\omega_{3}$ &   3.7253[1]  &   6.9862[1]  &   1.5063[2]  &   1.8758[2]  &   2.5527[2]\\ 
$\omega_{4}$ &   3.6068[1]  &   6.6665[1]  &   1.3683[2]  &   1.6778[2]  &   2.2011[2]\\ 
$\omega_{5}$ &   3.3715[1]  &   6.0587[1]  &   1.1414[2]  &   1.3662[2]  &    1.704[2]\\ 
$\omega_{6}$ &    3.008[1]  &    5.184[1]  &    8.768[1]  &   1.0236[2]  &   1.2209[2]\\ 
$\omega_{7}$ &   2.5525[1]  &   4.1845[1]  &   6.3831[1]  &    7.319[1]  &   8.5059[1]\\ 
$\omega_{8}$ &   2.0729[1]  &   3.2325[1]  &   4.5563[1]  &   5.1843[1]  &   5.9832[1]\\ 
$\omega_{9}$ &   1.6321[1]  &   2.4354[1]  &   3.2695[1]  &   3.7271[1]  &   4.3314[1]\\ 
$\omega_{10}$ &   1.2635[1]  &   1.8193[1]  &   2.3915[1]  &   2.7517[1]  &   3.2474[1]\\ 
$\omega_{11}$ &   9.7285  &   1.3632[1]  &   1.7939[1]  &   2.0939[1]  &   2.5192[1]\\ 
$\omega_{12}$ &   7.5094  &   1.0315[1]  &   1.3817[1]  &   1.6407[1]  &   2.0129[1]\\ 
$\omega_{13}$ &   5.8375  &   7.9115  &   1.0916[1]  &   1.3194[1]  &   1.6469[1]\\ 
$\omega_{14}$ &   4.5801  &   6.1592  &   8.8202  &   1.0842[1]  &   1.3718[1]\\ 
$\omega_{15}$ &   3.6298  &   4.8683  &   7.2655  &   9.0627  &   1.1576[1]\\ 
$\omega_{16}$ &   2.9053  &   3.9049  &   6.0798  &   7.6742  &   9.8569\\ 
$\omega_{17}$ &   2.3471  &   3.1759  &   5.1505  &   6.5592  &   8.4461\\ 
$\omega_{18}$ &   1.9124  &   2.6161  &   4.4038  &   5.6421  &   7.2685\\ 
$\omega_{19}$ &     1.57  &   2.1803  &     3.79  &   4.8731  &    6.274\\ 
$\omega_{20}$ &   1.2974  &    1.836  &   3.2758  &   4.2187  &   5.4271\\ 
$\omega_{21}$ &   1.0782  &   1.5604  &   2.8384  &   3.6559  &   4.7019\\ 
$\omega_{22}$ &   9.0039[-1]  &   1.3366  &   2.4616  &   3.1682  &    4.078\\ 
$\omega_{23}$ &    7.548[-1]  &   1.1527  &   2.1343  &   2.7435  &   3.5396\\ 
$\omega_{24}$ &   6.3468[-1]  &   9.9944[-1]  &   1.8481  &   2.3725  &   3.0734\\ 
$\omega_{25}$ &   5.3487[-1]  &   8.7016[-1]  &   1.5969  &   2.0478  &   2.6686\\ 
$\omega_{26}$ &   4.5139[-1]  &   7.5976[-1]  &   1.3759  &   1.7634  &   2.3161\\ 
$\omega_{27}$ &   3.8116[-1]  &   6.6437[-1]  &   1.1812  &   1.5144  &   2.0081\\ 
$\omega_{28}$ &   3.2178[-1]  &   5.8102[-1]  &   1.0098  &   1.2964  &   1.7382\\ 
$\omega_{29}$ &   2.7133[-1]  &   5.0746[-1]  &   8.5909[-1]  &   1.1059  &   1.5008\\ 
$\omega_{30}$ &   2.2829[-1]  &   4.4195[-1]  &   7.2678[-1]  &    9.397[-1]  &   1.2913\\ 
$\omega_{31}$ &   1.9145[-1]  &   3.8318[-1]  &   6.1095[-1]  &   7.9489[-1]  &   1.1058\\ 
$\omega_{32}$ &   1.5981[-1]  &   3.3015[-1]  &   5.0991[-1]  &   6.6906[-1]  &   9.4128[-1]\\ 
$\omega_{33}$ &   1.3258[-1]  &   2.8215[-1]  &   4.2213[-1]  &   5.5999[-1]  &   7.9502[-1]\\ 
$\omega_{34}$ &   1.0911[-1]  &   2.3865[-1]  &   3.4627[-1]  &    4.657[-1]  &   6.6503[-1]\\ 
$\omega_{35}$ &    8.888[-2]  &    1.993[-1]  &   2.8111[-1]  &   3.8439[-1]  &   5.4976[-1]\\ 
$\omega_{36}$ &   7.1473[-2]  &    1.639[-1]  &   2.2553[-1]  &   3.1446[-1]  &   4.4807[-1]\\ 
$\omega_{37}$ &   5.6557[-2]  &    1.323[-1]  &   1.7851[-1]  &   2.5446[-1]  &    3.591[-1]\\ 
$\omega_{38}$ &   4.3871[-2]  &   1.0445[-1]  &    1.391[-1]  &   2.0311[-1]  &   2.8221[-1]\\ 
$\omega_{39}$ &   3.3206[-2]  &   8.0308[-2]  &   1.0641[-1]  &    1.593[-1]  &   2.1681[-1]\\ 
$\omega_{40}$ &   2.4388[-2]  &   5.9819[-2]  &    7.961[-2]  &   1.2211[-1]  &   1.6226[-1]\\ 
$\omega_{41}$ &   1.7262[-2]  &   4.2895[-2]  &   5.7918[-2]  &   9.0832[-2]  &   1.1783[-1]\\ 
$\omega_{42}$ &   1.1673[-2]  &   2.9376[-2]  &   4.0634[-2]  &   6.4943[-2]  &    8.258[-2]\\ 
$\omega_{43}$ &   7.4569[-3]  &   1.9017[-2]  &   2.7154[-2]  &   4.4068[-2]  &    5.544[-2]\\ 
$\omega_{44}$ &   4.4315[-3]  &   1.1476[-2]  &   1.6976[-2]  &   2.7905[-2]  &   3.5224[-2]\\ 
$\omega_{45}$ &   2.3965[-3]  &   6.3285[-3]  &   9.6727[-3]  &   1.6113[-2]  &   2.0737[-2]\\ 
$\omega_{46}$ &   1.1397[-3]  &   3.0887[-3]  &    4.835[-3]  &   8.2051[-3]  &   1.0887[-2]\\ 
$\omega_{47}$ &   4.4978[-4]  &   1.2589[-3]  &   1.9961[-3]  &   3.4837[-3]  &   4.7639[-3]\\ 
$\omega_{48}$ &   1.3169[-4]  &   3.8074[-4]  &   6.0959[-4]  &   1.0995[-3]  &   1.5374[-3]\\ 
$\omega_{49}$ &   2.1915[-5]  &   6.4812[-5]  &    1.056[-4]  &   1.9486[-4]  &   2.7913[-4]\\ 
$\omega_{50}$ &   7.9107[-7]  &   2.3589[-6]  &   3.8988[-6]  &   7.3492[-6]  &   1.0706[-5]\\ 
\end{longtable}

\newpage

\setlength{\LTleft}{0pt}
\setlength{\LTright}{0pt} 
\footnotesize 
\begin{longtable}{@{\extracolsep\fill}llllll}
\caption{Dynamic 
electric-dipole polarizabilities of noble gas atoms for use with the 50-point Gauss-Legendre quadrature
method.\label{Tab:AlphaNob} }\\[-8pt]
  \caption*{\small{See page \pageref{sec.eot} for Explanation of Tables}}\\
\endhead\\
\hline
 \multicolumn{1}{l}{$\omega$} &  \multicolumn{1}{l}{He}  & 
   \multicolumn{1}{l}{Ne}  &  \multicolumn{1}{l}{Ar}  &
   \multicolumn{1}{l}{Kr}  &  \multicolumn{1}{l}{Xe} \\ 
\hline
0  &  1.383  &  2.669  &  11.08  &  16.79  &  27.16\\ 
\multicolumn{6}{c}{~}\\
$\omega_{1}$  &   1.3832  &   2.6693  &   1.1082[1]  &    1.679[1]  &   2.7156[1]\\ 
$\omega_{2}$  &   1.3831  &    2.669  &   1.1079[1]  &   1.6786[1]  &   2.7145[1]\\ 
$\omega_{3}$  &   1.3824  &   2.6678  &   1.1067[1]  &   1.6761[1]  &   2.7088[1]\\ 
$\omega_{4}$  &   1.3804  &   2.6641  &   1.1031[1]  &   1.6689[1]  &   2.6923[1]\\ 
$\omega_{5}$  &   1.3761  &    2.656  &   1.0954[1]  &   1.6534[1]  &   2.6571[1]\\ 
$\omega_{6}$  &   1.3681  &   2.6412  &   1.0814[1]  &   1.6256[1]  &   2.5947[1]\\ 
$\omega_{7}$  &    1.355  &    2.617  &   1.0591[1]  &   1.5819[1]  &   2.4986[1]\\ 
$\omega_{8}$  &   1.3355  &   2.5811  &   1.0269[1]  &   1.5202[1]  &    2.367[1]\\ 
$\omega_{9}$  &   1.3081  &   2.5315  &   9.8423  &   1.4405[1]  &   2.2034[1]\\ 
$\omega_{10}$  &   1.2721  &   2.4671  &   9.3171  &   1.3454[1]  &   2.0164[1]\\ 
$\omega_{11}$  &    1.227  &   2.3878  &   8.7097  &   1.2392[1]  &   1.8172[1]\\ 
$\omega_{12}$  &   1.1732  &   2.2948  &    8.044  &   1.1269[1]  &   1.6165[1]\\ 
$\omega_{13}$  &   1.1117  &     2.19  &   7.3471  &   1.0136[1]  &   1.4233[1]\\ 
$\omega_{14}$  &   1.0439  &   2.0762  &   6.6443  &   9.0318  &   1.2437[1]\\ 
$\omega_{15}$  &   9.7176[-1]  &   1.9562  &   5.9571  &   7.9877  &   1.0809[1]\\ 
$\omega_{16}$  &   8.9716[-1]  &   1.8327  &   5.3018  &   7.0221  &   9.3621\\ 
$\omega_{17}$  &   8.2198[-1]  &   1.7081  &   4.6893  &   6.1444  &   8.0938\\ 
$\omega_{18}$  &   7.4784[-1]  &   1.5844  &   4.1258  &   5.3569  &   6.9925\\ 
$\omega_{19}$  &   6.7605[-1]  &   1.4632  &   3.6139  &   4.6573  &   6.0419\\ 
$\omega_{20}$  &   6.0757[-1]  &   1.3456  &   3.1534  &   4.0403  &   5.2242\\ 
$\omega_{21}$  &   5.4303[-1]  &   1.2325  &   2.7424  &   3.4989  &   4.5216\\ 
$\omega_{22}$  &   4.8284[-1]  &   1.1244  &   2.3777  &   3.0256  &   3.9179\\ 
$\omega_{23}$  &   4.2714[-1]  &   1.0215  &   2.0555  &   2.6129  &   3.3983\\ 
$\omega_{24}$  &   3.7596[-1]  &   9.2402[-1]  &   1.7719  &   2.2535  &   2.9499\\ 
$\omega_{25}$  &   3.2919[-1]  &   8.3201[-1]  &   1.5229  &   1.9409  &   2.5615\\ 
$\omega_{26}$  &   2.8667[-1]  &   7.4541[-1]  &   1.3049  &   1.6691  &   2.2239\\ 
$\omega_{27}$  &   2.4818[-1]  &   6.6416[-1]  &   1.1142  &   1.4329  &   1.9291\\ 
$\omega_{28}$  &    2.135[-1]  &   5.8814[-1]  &   9.4788[-1]  &   1.2276  &   1.6705\\ 
$\omega_{29}$  &   1.8237[-1]  &   5.1726[-1]  &   8.0296[-1]  &   1.0493  &   1.4427\\ 
$\omega_{30}$  &   1.5457[-1]  &   4.5142[-1]  &   6.7699[-1]  &   8.9435[-1]  &   1.2414\\ 
$\omega_{31}$  &   1.2987[-1]  &   3.9052[-1]  &   5.6773[-1]  &   7.5977[-1]  &   1.0627\\ 
$\omega_{32}$  &   1.0804[-1]  &    3.345[-1]  &   4.7325[-1]  &   6.4289[-1]  &   9.0391[-1]\\ 
$\omega_{33}$  &   8.8889[-2]  &    2.833[-1]  &   3.9181[-1]  &   5.4141[-1]  &   7.6262[-1]\\ 
$\omega_{34}$  &   7.2211[-2]  &   2.3685[-1]  &   3.2189[-1]  &   4.5327[-1]  &   6.3702[-1]\\ 
$\omega_{35}$  &   5.7821[-2]  &   1.9512[-1]  &   2.6211[-1]  &   3.7666[-1]  &   5.2567[-1]\\ 
$\omega_{36}$  &   4.5539[-2]  &   1.5804[-1]  &   2.1123[-1]  &   3.1002[-1]  &   4.2747[-1]\\ 
$\omega_{37}$  &   3.5189[-2]  &   1.2553[-1]  &   1.6814[-1]  &   2.5197[-1]  &   3.4156[-1]\\ 
$\omega_{38}$  &   2.6598[-2]  &   9.7463[-2]  &   1.3181[-1]  &   2.0141[-1]  &    2.673[-1]\\ 
$\omega_{39}$  &   1.9593[-2]  &   7.3682[-2]  &   1.0136[-1]  &   1.5751[-1]  &   2.0416[-1]\\ 
$\omega_{40}$  &   1.4001[-2]  &   5.3982[-2]  &   7.6035[-2]  &   1.1976[-1]  &   1.5163[-1]\\ 
$\omega_{41}$  &    9.648[-3]  &   3.8104[-2]  &   5.5247[-2]  &   8.7897[-2]  &   1.0912[-1]\\ 
$\omega_{42}$  &   6.3624[-3]  &    2.573[-2]  &   3.8541[-2]  &   6.1803[-2]  &   7.5839[-2]\\ 
$\omega_{43}$  &   3.9735[-3]  &   1.6472[-2]  &   2.5547[-2]  &   4.1315[-2]  &   5.0763[-2]\\ 
$\omega_{44}$  &    2.316[-3]  &   9.8772[-3]  &   1.5877[-2]  &   2.6037[-2]  &   3.2591[-2]\\ 
$\omega_{45}$  &   1.2329[-3]  &   5.4433[-3]  &   9.0623[-3]  &   1.5259[-2]  &   1.9829[-2]\\ 
$\omega_{46}$  &    5.795[-4]  &   2.6642[-3]  &   4.5722[-3]  &   8.0494[-3]  &   1.1001[-2]\\ 
$\omega_{47}$  &   2.2691[-4]  &   1.0845[-3]  &      1.9[-3]  &   3.5374[-3]  &    5.073[-3]\\ 
$\omega_{48}$  &    6.614[-5]  &   3.2458[-4]  &   5.7452[-4]  &   1.1184[-3]  &   1.6545[-3]\\ 
$\omega_{49}$  &   1.0985[-5]  &    5.458[-5]  &   9.7061[-5]  &   1.9243[-4]  &   2.8817[-4]\\ 
$\omega_{50}$  &   3.9634[-7]  &   1.9807[-6]  &   3.5573[-6]  &   7.0524[-6]  &   1.0532[-5]\\ 
\end{longtable}

\newpage




\end{document}